\documentclass[twocolumn,prb,showpacs,preprintnumbers,amsmath,amssymb]{revtex4}

\usepackage{graphicx}
\usepackage{dcolumn}
\usepackage{bm}

\begin{document}

\preprint{APS/123-QED}

\title{Energy gap formation in a valence fluctuating compound CeIrSb \\ probed by Sb NMR and NQR}

\author{Y.~Kawasaki}
\author{M.~Izumi}%
\author{Y.~Kishimoto}%
\author{T.~Ohno}%
\affiliation{Institute of Technology and Science, The University of Tokushima, Tokushima 770-8506, Japan}
\author{H.~Tou}%
\author{Y.~Inaoka}%
\author{M.~Sera}%
\author{K.~Shigetoh}%
\affiliation{ADSM, Hiroshima University, Higashi-Hiroshima 739-8530, Japan}
\author{T.~Takabatake}%
\affiliation{ADSM, Hiroshima University, Higashi-Hiroshima 739-8530, Japan}
\affiliation{IAMR, Hiroshima University, Higashi-Hiroshima 739-8530, Japan}

\date{\today}

\begin{abstract}
Sb-NMR/NQR study has revealed a formation of a pseudogap at the Fermi level in the density of states in a valence fluctuating compound CeIrSb\@.
The nuclear spin-lattice relaxation rate divided by temperature, $1/T_1T$ has a maximum around 300 K and decreases significantly as $1/T_1T \sim T^2$, followed by a $1/T_1T$ = const.\ relation at low temperature.
This temperature dependence of $1/T_1T$ is well reproduced by assuming a V-shaped energy gap with a residual density of states at the Fermi level.
The size of energy gap for CeIrSb is estimated to be about 350 K, which is by one order of magnitude larger than those for the isostructural Kondo semiconductors CeRhSb and CeNiSn\@.
Despite the large difference in the size of energy gap, CeIrSb, CeRhSb and CeNiSn are indicated to be classified into the same group revealing a V-shaped gap due to c-f hybridization.
The temperature dependence of the Knight shift measured in a high magnetic field agrees with the formation of this pseudogap.
\end{abstract}

\maketitle
\sloppy

\section{Introduction}

A kind of rare-earth compounds called Kondo insulator or semiconductor undergoes a crossover from a metallic state into an insulating ground one with decreasing temperature, where a narrow gap or a pseudogap is formed at the Fermi level ($E_{\rm F}$) in a coherent heavy fermion band.\cite{takabatake98,fisk95}
These energy gap openings are generally considered to be derived from c-f hybridization between a wide conduction band and a 4f band that is renormalized as a heavy fermion band at $E_{\rm F}$ at low temperature.
The smallest gap Kondo insulators, CeRhSb and CeNiSn with orthorhombic structure, are exotic among them: they appear to develop an anisotropic energy gap, while other systems such as Ce$_3$Bi$_4$Pt$_3$ and YbB$_{12}$ with cubic structure have a well-defined size of energy gap.
The anisotropic energy gap has been suggested by NMR in CeRhSb and CeNiSn for the first time, where the power law of nuclear spin-lattice relaxation rate $1/T_1$ in its temperature dependence is explained by a narrow V-shaped pseudogap at $E_{\rm F}$ in the density of states (DOS).\cite{kyogaku90,nakamura94}
These results, together with transport,\cite{takabatake94} specific heat,\cite{nishigori96} and tunneling spectroscopy measurements,\cite{ekino95} indicate the formation of a new type of semimetal with an anisotropic gap for CeRhSb and CeNiSn\@.
Such an anisotropic gap is proposed to originate from an anisotropic hybridization of conduction band with a certain ground state of 4f electron in the crystalline field \cite{ikeda96,moreno00} and, therefore, it may be closely related to its crystal lattice.

On the other hand, it has been reported that CeRhAs shows a gap opening over the entire Fermi surface by various measurements,\cite{yoshii96,shimada02,matsumura03,adroja06} although CeRhAs is isostructural to CeRhSb and CeNiSn at room temperature.
Recent investigation in CeRhAs indicate that successive structural modulations at low temperature are closely related to its gap property.\cite{sasakawa02,sasakawa05a,matsunami02,ogita03}
Thus, the detailed mechanism of the anisotropic gap formation has received new attention in connection with a crystal lattice.

Recently, another isostructural compound CeIrSb with orthorhombic $\epsilon$-TiNiSi type structure was synthesized.\cite{hasse02, sasakawa05}
The lattice parameters for CeIrSb are $a =$ 7.351, $b =$ 4.5744, and $c =$ 7.935 \AA.
On going from CeRhSb to CeIrSb, the $a$ and $b$ parameters decrease by 0.9\% but the $c$ parameter increases by a similar amount, thus the unit cell volume is decreased by 0.8\%.
This anisotropic shrink in the lattice may strengthen c-f hybridization.
Actually, the large negative Curie-Weiss temperature of $-1300$ K and the gradual decrease in resistivity with decreasing temperature suggest that CeIrSb is a valence fluctuating compound with stronger c-f hybridization than that for CeRhSb.\cite{sasakawa05}
The value of $C/T$ below 10 K is less than that of CeRhSb, which proves that the DOS at $E_{\rm F}$ is strongly reduced.
It is notable that the resistivity of CeIrSb increases below 8 K, which is reminiscent of the similar behavior of resistivity in CeRhSb\@.
Although this upturn in the resistivity may be associated with an energy gap opening, thermopower does not show an enhanced value at low temperature, which disagrees with a distinct characteristic of narrow gap semiconductors.\cite{hundley94, takabatake94, iga01}
Thus, it has been still controversial whether the ground state of CeIrSb is a gapped one or not.
In order to make this point clear from a microscopic viewpoint, we have performed Sb-NMR/NQR measurements on CeIrSb and its lanthanum analog LaIrSb\@.

\section{Experimental aspect}

Detailed preparation method of the polycrystalline samples was reported in the literature.\cite{sasakawa05}
X-ray powder diffraction (XRD) and electron-probe microanalysis (EPMA) detected impurity phases of CeSb and CeIr$_2$ with several percent.\cite{sasakawa05}
The amounts of impurity phases of CeSb and CeIr$_2$ were estimated to be 5\% and 3\%, respectively, from the combined analysis of XRD, EPMA and specific-heat anomaly due to the antiferromagnetic order of CeSb at 17 K\@.\cite{hulliger75}
NMR/NQR method is, however, a very local probe and therefore yields information on the majority phase, CeIrSb\@.
In this meaning, physical quantities obtained by NMR/NQR method have its inherent advantage over macroscopic ones that are affected by the magnetic impurity CeSb, for example.\cite{sasakawa05}
We grounded these polycrystals into grains with smaller size than 100 $\mu$m for NMR/NQR measurement.
The measurements were made at temperatures between 1.6 K and 300 K by employing a phase-coherent pulsed NMR/NQR spectrometer.
NMR (NQR) spectrum was measured by tracing the integrated spin-echo signal as the function of an external magnetic field (the frequency of pulsed rf field in zero external magnetic field).
We obtained nuclear spin-lattice relaxation rate, $1/T_1$ by fitting the longitudinal nuclear magnetization recovery after saturating pulse to theoretical functions given in Refs.\ [24] and [25].
The high quality of the fittings with single component of $1/T_1$ guarantees that obtained $1/T_1$'s will make clear intrinsic nature of CeIrSb, unaffected by any interruption from the impurity phases.\cite{kawasaki07}

\begin{figure}[tbp]
\includegraphics[scale=0.5]{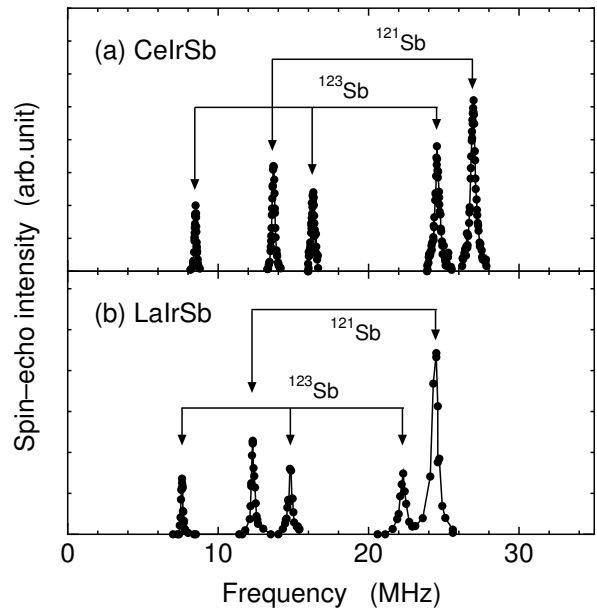}
\caption{$^{121, 123}$Sb-NQR spectra of (a) CeIrSb and (b) LaIrSb at 4.2 K\@.}
\end{figure}

\section{Results and discussion}

\subsection{Nuclear quadrupole resonance and spin-lattice relaxation}

Figure 1 displays the $^{121,123}$Sb-NQR spectra of (a) CeIrSb and (b) LaIrSb for two Sb isotopes at 4.2 K\@.$^{121}$Sb ($^{123}$Sb) has a nuclear spin $I = 5/2$ (7/2) and therefore exhibits two (three) NQR transitions.
In CeIrSb, the resonance lines for $^{121}$Sb ($^{123}$Sb) are found at 13.68 and 26.95 MHz (8.50, 16.30 and 24.56 MHz), corresponding to $^{121}\nu_Q = 13.51$ MHz ($^{123}\nu_Q = 8.20$ MHz) with $\eta$ = 0.107.
Here, the nuclear quadrupole frequency $\nu_Q$ and the asymmetry parameter $\eta$ are defined as $\nu_Q=\nu_z=\frac{3e^2Q}{2I(2I-1)h}\frac{\partial V}{\partial z}$ and $\eta = |\nu_x-\nu_y|/\nu_z$, respectively, with the nuclear quadrupolar moment $Q$ and the electric field gradient at the position of the nucleus $\partial V/\partial \alpha$ ($\alpha = x, y, z$).
Since no change in the spectral shape across 17 K, at which CeSb exhibits the magnetic order,\cite{kawasaki07} has been observed, we consider these signals originate from CeIrSb\@.
Any other signal from impurity phase is not observed in the NQR spectra, presumably because these signal intensities are below the sensitivity of our spectrometer.
The line width for the transition $\pm1/2 \leftrightarrow \pm3/2$ of $^{123}$Sb in CeIrSb is 89 kHz, which is about four times larger than 23 kHz for the corresponding transition observed in CeRhSb,\cite{nakamura94} but smaller than 200-300 kHz observed for CeRhAs.\cite{matsumura03}
As for LaIrSb, the resonance lines for $^{121}$Sb ($^{123}$Sb) are found at 12.29 and 24.43 MHz (7.59, 14.79 and 22.27 MHz), corresponding to $^{121}\nu_Q = 12.22$ MHz ($^{123}\nu_Q = 7.43$ MHz) with $\eta$ = 0.078.

Figure 2 displays the temperature dependence of $1/T_1T$ for CeIrSb (closed circles) with that for CeRhSb (crosses [4]) in zero field.
Here, these $1/T_1$'s were measured at the resonance line arising from the $\pm3/2 \leftrightarrow \pm5/2$ transitions of $^{121}$Sb for both compounds.
The value of $^{121}(1/T_1)/^{123}(1/T_1) =$ 2.9 at 4.2 K for CeIrSb, is comparable to $(^{121}\gamma_{\rm n}/^{123}\gamma_{\rm n})^2$ = 3.4 expected when magnetic interaction is dominantly responsible for the nuclear spin relaxation.
Here, $^{\beta}(1/T_1)$ and $^{\beta}\gamma_{\rm n}$ are the relaxation rate and the nuclear gyromagnetic ratio of Sb isotopes with its mass number $\beta$.
This fact proves that the leading contribution to the nuclear spin relaxation arises from magnetic interactions.

\begin{figure}[tbp]
\includegraphics[scale=0.43]{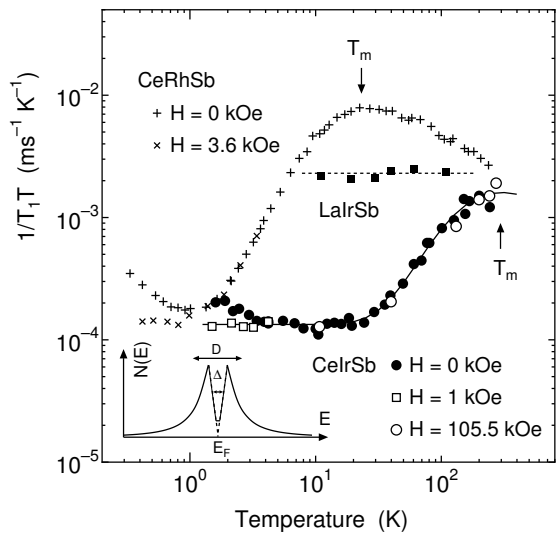}
\caption{\protect Temperature dependences of $1/T_1T$ for CeIrSb (closed circles), LaIrSb (closed squares) and CeRhSb (crosses \ [4]) in zero field.
The open squares and circles represent $1/T_1T$ for CeIrSb in magnetic fields of 1 and 105.5 kOe, respectively.
The solid line indicates the calculation by assuming an effective DOS with V-shaped gap structure at $E_{\rm F}$ as drawn in the bottom left.}
\end{figure}

The overall temperature dependence of $1/T_1T$ for CeIrSb is qualitatively similar to that for CeRhSb\@.
$1/T_1T$ for CeIrSb has a maximum around $T_{\rm m}$ = 300 K and decreases significantly as $1/T_1T \sim T^2$ below this temperature, followed by a weak upturn below 4 K\@.
This decrease in $1/T_1T$ with decreasing temperature indicates that CeIrSb has an energy gap at $E_{\rm F}$ in the DOS at low temperature.
The similarity in $1/T_1$ is demonstrated in Fig.\ 3, where the normalized relaxation rate $(1/T_1)/(1/T_1)_{\rm m}$ is plotted against the normalized temperature $T/T_{\rm m}$ for CeIrSb and CeRhSb\@.
Here, $T_{\rm m}$ and $(1/T_1)_{\rm m}$ are the temperature at which $1/T_1T$ has a maximum and $1/T_1$ at $T_{\rm m}$ for each compound, respectively.
The temperature dependences of $1/T_1$ for both compounds are scaled with $T_{\rm m}$ except those in low temperature region.
This deviation from the scaling at low temperature is explained by the difference in the amount of residual DOS at $E_{\rm F}$ as discussed later.
It should be noted that this scaling between $1/T_1$ and $T_{\rm m}$ is good although the difference in $T_{\rm m}$'s for CeIrSb and CeRhSb is as large as 10 times.
This result indicates that these two compounds have a similar gap structure and that the magnitudes of the energy gap are scaled with $T_{\rm m}$\@.

The weak upturn in $1/T_1T$ below 4 K for CeIrSb is probably due to the relaxation process by spin fluctuations of paramagnetic Ce spins through direct mutual spin flipping between electron and nuclear spins, which has been also observed in CeRhSb.\cite{nakamura94}
In order to suppress this extrinsic relaxation channel, we have measured $1/T_1T$ in a small magnetic field of $H$ = 1 kOe, because the flipping may be prevented by a magnetic field which leads to a different Zeeman splitting between electron and nuclear spin levels.
As a matter of fact, the upturn is suppressed and $1/T_1T$ stays constant down to 1.6 K in 1 kOe as indicated by open squares in Fig.~2\@.
Similar suppression has been also observed in CeRhSb \cite{nakamura94} by applying $H = 3.6$ kOe as shown in Fig.~2\@.
In small magnetic fields, where the extrinsic relaxation contribution is suppressed, $1/T_1T$ for CeIrSb is independent of temperature below 20 K.
This result indicates the existence of residual DOS at $E_{\rm F}$ even though a V-shaped gap is made.

\begin{figure}[tbp]
\includegraphics[scale=0.5]{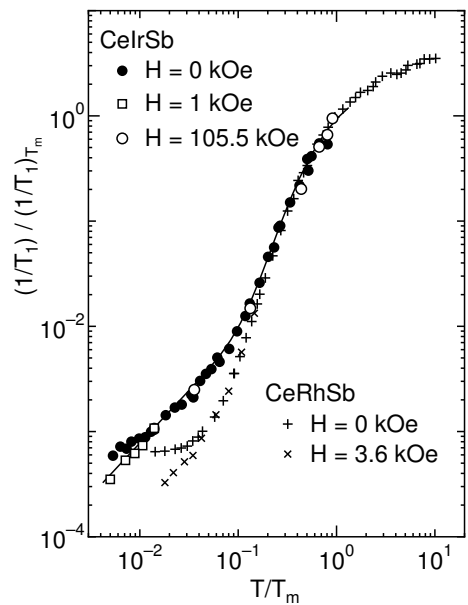}
\caption{\protect $1/T_1$ normalized by the value at $T_{\rm m}$ as a function of the normalized temperature $T/T_{\rm m}$  for CeIrSb (closed circles) and CeRhSb (crosses [4]) in zero field.
The open squares and circles represent the data for CeIrSb at $H =$ 1 and 105.5 kOe, respectively.
The solid curve gives the reproduction by the best fitted V-shaped gap.
}
\end{figure}

Since the temperature dependences of $1/T_1T$ for CeIrSb and CeRhSb are quite similar, the V-shaped gap model used in CeRhSb \cite{nakamura94} is applicable to the gapped state in CeIrSb\@.
Namely, we assume an effective DOS, $N(E)$ having a V-shaped gap structure with a residual DOS at $E_{\rm F}$ as shown in the bottom left of Fig.~2\@.
The temperature dependence of $1/T_1T$ in CeIrSb could be well fitted to the relation
\begin{equation}
1/T_1T\propto \int\!\!N(E)^2\left\{-\frac{\partial f(E)}{\partial E}\right\}dE,
\label{eq:1}
\end{equation}
where $f(E)$ is the Fermi distribution function, with the band width $D$ = 1800 K, the gap $\Delta$ = 350 K and the fraction of residual DOS against a value without gap at $E_{\rm F}$ for the Lorentzian band $N(E_{\rm F})_{\rm res}/N_0(E_{\rm F})$ = 0.193.
The result of calculation, indicated by solid lines in Figs.~2 and 3, is in good agreement with the experimental data when the above-mentioned extrinsic relaxation channel is suppressed in a small magnetic field.

These parameters for CeIrSb as well as those for CeRhSb \cite{nakamura94} and CeNiSn \cite{nakamura94} are listed in Table I.\@
The values of $D$ and $\Delta$ for CeIrSb are much larger than those for CeRhSb and CeNiSn, respectively.
The larger value of $D$ for CeIrSb is consistent with the weaker temperature dependence of the magnetic susceptibility, suggesting a stronger c-f hybridization.\cite{sasakawa05}
Despite the values of $D$ and $\Delta$ for CeIrSb by one order of magnitude larger than those for others, it is notable that $\Delta$ is scaled with $D$ among these compounds, which is expected when the c-f hybridization is essential for the V-shaped gap formation.
In this meaning, it is considered that these compounds are classified into the same group revealing a V-shaped gap due to c-f hybridization and that the much larger band width for CeIrSb brings about the much larger magnitude of the energy gap.
The larger value of $N(E_{\rm F})_{\rm res}/N_0(E_{\rm F})$ for CeIrSb than those for CeRhSb and CeNiSn may originate from a possible off-stoichiometry in composition or effective carrier doping effect,\cite{nakamura96} which may be associated with the larger NQR line width of 89 kHz for CeIrSb than that of 23 kHz for CeRhSb\@.

LaIrSb, a metallic and non-magnetic analog of CeIrSb, may offer information of conduction electron states without 4f electrons.
As indicated by closed squares in Fig.~2, $1/T_1T$ of LaIrSb is almost independent of temperature, characteristics of a normal metallic state, where a DOS near $E_{\rm F}$ is finite and almost flat with a temperature-independent band structure.
In such a metallic state, we may consider $1/T_1T \propto N(E_{\rm F})^2$ as approximately obtained from Eq.~(\ref{eq:1})\@.
The value of $1/T_1T$ below 10 K in CeIrSb, 0.134 s$^{-1}$K$^{-1}$, is by one order of magnitude smaller than that of LaIrSb, 2.32 s$^{-1}$K$^{-1}$, although a lanthanum has no 4f electrons.
Therefore, if we assume a comparable hyperfine coupling constant for CeIrSb and LaIrSb, the smaller value of $1/T_1T$ in CeIrSb is considered to be related to the reduction in $N(E_{\rm F})$ of CeIrSb at low temperature due to the V-shaped energy gap.
The ratio of the residual DOS for CeIrSb against that for LaIrSb is estimated to be $\sqrt{(1/T_1T)_{\rm CeIrSb}/(1/T_1T)_{\rm LaIrSb}}= \sqrt{0.134/2.31} = 0.241$, about 25\% larger than $N(E_{\rm F})_{\rm res}/N_0(E_{\rm F}) = 0.193$ obtained in the previous paragraph.
A possible contribution of 4f electrons to the conduction band in CeIrSb due to c-f hybridization may be responsible for this difference.

\begin{table}
\caption{\protect
List of quasi-particle band width $D$, energy gap $\Delta$ and the fraction of residual DOS against a value without gap at $E_{\rm F}$ for the Lorentzian band $N(E_{\rm F})_{\rm res}/N_0(E_{\rm F})$ for CeIrSb, CeRhSb \cite{nakamura94} and CeNiSn.\cite{nakamura94}
} 
\begin{center}
\begin{tabular}{p{3cm}p{1.7cm}p{1.7cm}p{1.7cm}}\hline
Samples & CeIrSb & CeRhSb  & CeNiSn \\ \hline
Band width $D$ (K) & 1800 & 210 & 140  \\
Energy gap $\Delta$ (K) & 350 & 28 & 14 \\
$N(E_{\rm F})_{\rm res}/N_0(E_{\rm F})$ & 0.193 & 0.085 & 0.077 \\ \hline
\end{tabular}
\end{center}
\end{table}

\subsection{Nuclear magnetic resonance and Knight shift}

To gain further insight into the gap property of CeIrSb, we have performed $^{121}$Sb-NMR measurements in high magnetic field.
The field-swept $^{121}$Sb-NMR spectrum measured at fixed frequency of 107.3 MHz at 10 K is shown in the inset of Fig.~4\@.
It indicates a distribution of resonance field at Sb site characterized by six peaks at 9.34, 9.84, 10.39, 10.58, 11.00 and 11.69 T in the polycrystalline sample, so-called powder pattern for nuclei with $I = 5/2$.
Here, a second order perturbation of nuclear quadrupole interaction added to Zeeman interaction yields a splitting of the central line into two peaks.
Neither broadening nor splitting in the resonance lines has been observed down to 5 K (not shown), indicating that CeIrSb is non magnetic at the ground state and that these resonance lines come from CeIrSb\@.
Small peaks at 9.69 T and 11.10 T may come from CeSb\@.

\begin{figure}[tbp]
\includegraphics[scale=0.5]{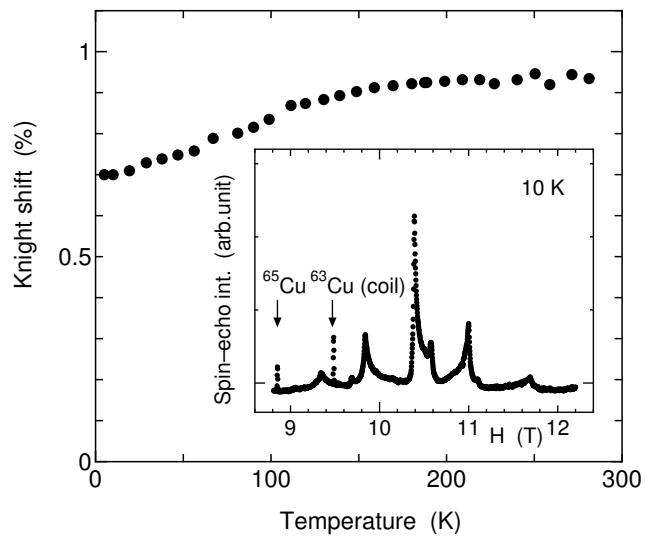}
\caption{Temperature dependence of $^{121}$Sb Knight shift for CeIrSb\@. Inset shows the field-swept $^{121}$Sb-NMR spectrum at 10 K measured with the fixed rf frequency of 107.3 MHz\@.}
\end{figure}

We show the temperature dependences of $1/T_1T$ and $(1/T_1)/(1/T_1)_{\rm m}$ measured at the central resonance lines by open circles in Figs.~2 and 3, respectively.
Apparently, these relaxation rates in high magnetic field are almost identical to those in zero and low field, indicating that the gap structure is not affected by the applied magnetic field at least up to 10 T\@.
It is a remarkable contrast to the case for CeNiSn, in which the pseudogap is known to be progressively suppressed with increasing the applied magnetic field higher than 2 T, while no dramatic change was found below 2 T\@.\cite{takabatake98}
These facts have been interpreted as a result of the presence of a flat part at the bottom of the pseudogap and the respective Zeeman shift of valence up-spin and conduction down-spin bands which induces the overlapping of these bands at high magnetic field.\cite{izawa96,nakamura96a}
The robust gap unaffected by the external magnetic field in CeIrSb is consistent with the much larger sizes of energy gap and residual DOS than those for CeNiSn\@.

In the main panel of Fig.\ 4, we show the temperature dependence of the Knight shift, $K(T)$, which is obtained from the two peaks of the $-1/2 \leftrightarrow 1/2$ transition with taking non-zero value of $\eta$ = 0.107 into account.
It is notable that $K(T)$ reflects the intrinsic magnetic property of CeIrSb, while $\chi(T)$ is largely affected by the impurity phase CeSb giving rise to the Curie-Weiss-like upturn below 250 K.\cite{sasakawa05}
In general, $K(T)$ is dominated by two contributions, a temperature dependent spin part $K_s(T)$ and a temperature independent Van Vleck one $K_{\rm VV}$, as $K(T) = K_s(T)+K_{\rm VV}$\@.
The temperature dependent portion of $K(T)$ arises from $K_s(T)$ part, which is proportional to a corresponding spin susceptibility $\chi_s(T)$ via a dominant transferred hyperfine coupling.
Therefore, the very weak temperature dependence in $K(T)$ above 200 K suggests that this compound is in the regime of intermediate valence systems.
The decrease in $K(T)$ below 200 K is considered to be associated with the decrease in the DOS at $E_{\rm F}$ in CeIrSb\@.
At the lowest temperature, there remains large $K(T)$ of 0.7 \%, which may be attributed to the large residual DOS and the Van Vleck contribution.
By subtracting $K_s$ arising from the residual DOS, $N(E_{\rm F})_{\rm res}/N_0(E_{\rm F}) = 0.194$, at low temperature, the Van Vleck contribution is estimated to be 0.65 \%.
Such a large Van Vleck contribution has been also observed in other Kondo insulators and semiconductors \cite{takabatake98} and explained in a strongly correlated regime, where the Van Vleck susceptibility is enhanced by the same renormalization factor as for the effective mass.\cite{kontani96}
Although the 4f electronic state of CeIrSb is not expected to be highly renormalized with the conduction band as in a heavy fermion material, some renormalization effects might enhance $K_{\rm VV}$ and yield the large value of $K(T)$ at low temperature.

\section{Conclusion}
The valence fluctuating compound CeIrSb and its La analog LaIrSb have been investigated by Sb-NMR/NQR measurements.
The $1/T_1T$ for CeIrSb has a maximum around 300 K and decreases significantly as $1/T_1T \sim T^2$, followed by a $T_1T$ = const.\ relation at low temperature.
This temperature dependence of $1/T_1T$ indicates a V-shaped energy gap with a residual DOS at $E_{\rm F}$\@.
The size of energy gap for CeIrSb is estimated to be about 350 K, which is by one order of magnitude larger than the respective 28 K and 14 K for CeRhSb and CeNiSn\@.
Despite the very large difference in the size of energy gap, the temperature dependence of $1/T_1$ has revealed the scaling between the band width and the magnitude of energy gap among these compounds.
This fact indicates that these compounds are classified into the same group exhibiting a V-shaped gap due to c-f hybridization.
The temperature dependence of the Knight shift for CeIrSb measured in a high magnetic field is consistent with the formation of the pseudogap at $E_{\rm F}$\@.

\end{document}